\documentstyle[aps,preprint,epsfig]{revtex}
\tightenlines
\begin{document}
\preprint{\parbox{6cm}{\flushright CLNS 99/1632}}
\newcommand{\bea}{\begin{eqnarray}}
\newcommand{\eea}{\end{eqnarray}}
\newcommand{\beq}{\begin{equation}}
\newcommand{\eeq}{\end{equation}}
\newcommand{\bay}{\begin{array}}
\newcommand{\eay}{\end{array}}
\title{Rescattering and Electroweak Penguin Effects
in Determinations of the Weak Phase $\gamma$
\footnote{Invited talk given at the Beauty '99 conference, 
June 21-25, Bled, Slovenia.}}
\author{Dan Pirjol\footnote{pirjol@mail.lns.cornell.edu}}
\address{Floyd R. Newman Laboratory for Nuclear Studies,
Cornell University, Ithaca, New York 14853}
\date{\today}
\maketitle
 
\begin{abstract}
Determinations of the CKM phase $\gamma$ from weak
nonleptonic $B$ decays are affected by electroweak (EW) penguins
and rescattering effects. In this talk I explain how the EW penguin effects can be
controlled with the help of SU(3) symmetry, by relating them to tree-level
amplitudes. The impact of the final-state interactions on the
determination of $\gamma$ from $B^+\to K\pi$ decays is studied
numerically, showing that they can be important. A few alternative methods
are discussed which use additional decays to eliminate their effects.
\end{abstract}

\narrowtext

\section{Introduction}

The Cabibbo-Kobayashi-Maskawa (CKM) mixing matrix is one of the most important free 
parameters of the Standard Model, encoding CP violation as complex phases in some 
of its matrix elements. These effects manifest themselves only in the couplings 
of the third generation, which makes the weak decays of bottom flavored hadrons an
ideal setting for their study. Preliminary results from CLEO and CDF are already providing 
tantalizing evidence for nontrivial phases in some of the CKM matrix elements,
and more precise determinations are expected soon from the $B$ factories presently
being commisioned. Ultimately, studies of weak $B$ decays will help test the Standard
Model mechanism of CP violation and explore the possible existence of new physics.

For our purposes, the following approximate form of the CKM matrix given by
Wolfenstein will be sufficient
\bea
\left( \begin{array}{ccc}
V_{ud} & V_{us} & V_{ub} \\
V_{cd} & V_{cs} & V_{cb} \\
V_{td} & V_{ts} & V_{tb} 
\end{array}\right) 
 \simeq 
\left( \begin{array}{ccc}
1-\lambda^2/2 & \lambda & A\lambda^3 (\rho-i\eta) \\
-\lambda & 1-\lambda^2/2 & A\lambda^2 \\
A\lambda^3(1-\rho-i\eta) & -A\lambda^2 & 1
\end{array}\right)\,.
\eea
In this parametrization the CP-violating phases are restricted to two
matrix elements $V_{ub}=A\lambda^3 R_b e^{-i\gamma}$ and $V_{td}=A\lambda^3
R_t e^{-i\beta}$, where we defined $R_b=\sqrt{\rho^2+\eta^2}$ and 
$R_t=\sqrt{(1-\rho)^2+\eta^2}$. Finally, a third weak phase $\alpha$ is defined by
$\alpha+\beta+\gamma=\pi$. The three weak phases are identical with the angles 
of the unitarity triangle following from the unitarity condition of the CKM matrix
\beq\label{unittriangle}
V_{ud} V^*_{ub} + V_{cd} V^*_{cb} + V_{td} V^*_{tb} =0\,.
\eeq

Numerically, the best known parameters are $\lambda \simeq |V_{us}| = 0.2196\pm 0.0023$ 
and $A\equiv |V_{cb}|/\lambda^2 = 0.819\pm 0.058$ \cite{PDG}. The remaining parameters
have been estimated from global fits of the unitarity triangle as $R_b = |V_{ub}/
V_{cb}|/\lambda = 0.41\pm 0.07$ and $R_t = |V_{td}/V_{cb}|/\lambda = 1.01\pm 0.21$
\cite{Ros}. Although knowledge of the sides of the triangle (\ref{unittriangle}) 
is sufficient to determine its angles too, one would like to measure the latter
directly, which would provide a consistency test of the whole picture.

Several methods have been proposed for determining the weak phases from $B$ decay
data, the most popular of which can be divided into two large classes: a) methods using 
mixing-induced CP
violation in neutral ($B_d$ or $B_s$) decays to CP eigenstates and b) methods using 
time-independent charged and/or neutral $B$ decay rates (for a discussion of other 
methods see the contribution by R. Fleischer
in these proceedings). The best known methods of type a) include the determination
of the weak phase $\alpha$ from $B^0(t)\to \pi^+\pi^-$ decays and of the phase
$\beta$ from $B^0(t)\to J/\psi K_S$ \cite{BiSa}. Such methods are more demanding
from a practical point of view, as they require
time-dependent measurements of the CP asymmetry.

The second class of methods employs the approximate flavour SU(3) symmetry
of the strong interactions \cite{SU(3)1,GHLR,SU(3)3}. 
The basic idea is that any $B$ decay amplitude is given
by a linear combination of (unknown) strong matrix amplitudes $T_j$ times CKM 
factors $\xi_j$ as $A = \sum_j \xi_{j} T_j$.
The strong amplitudes $T_j$ corresponding to different decays are related by 
SU(3) symmetry, such that one aims to eliminate them completely by combining
sufficiently many physical decay amplitudes, in order to determine the $\xi_j$ factors.
A particularly elegant version of this approach is formulated in a graphical language,
wherein the weak phases appear as angles in diagrams constructed from physical decay
amplitudes \cite{GHLR,GL,GW}.
While simple and attractive from an experimental point of view, this type of
methods are fraught
with theoretical uncertainties such as SU(3) breaking effects, final state interactions
and electroweak penguin effects. We will discuss these issues at length in the
following sections.

\section{SU(3) flavor symmetry and nonleptonic $B$ decays}

The flavour symmetry of the strong interactions plays an useful role in
organizing the structure of weak decay amplitudes of $B$ mesons into two
pseudoscalars. The effective weak nonleptonic Hamiltonian responsible for these
decays is given by
\bea\nonumber
{\cal H} &=& \frac{G_F}{\sqrt2}\sum_{q=d,s}\left(
\sum_{q'=u,c}
V_{q'b}^* V_{q'q} [ c_1 (\bar bq')_{V-A}(\bar q'q)_{V-A} + 
c_2 (\bar bq)_{V-A}(\bar q'q')_{V-A}] \right.\\\label{Ham}
& &\left.\qquad\qquad - V_{tb}^* V_{tq}\sum_{i=3}^{10} c_i Q_i^{(q)}\right)\,,
\eea
where the eight operators $Q_i^{(q)}$ include four QCD penguin-type and
four EW penguin-type operators
\bea\label{Q3}
Q_3^{(q)} = (\bar bq)_{V-A}\sum_{q'=u,d,s,c}(\bar q'q')_{V-A}\,,\,\,\,
Q_4^{(q)} = (\bar b_i q_j)_{V-A}\sum_{q'=u,d,s,c}(\bar q'_j q'_i)_{V-A}& &\\
Q_5^{(q)} = (\bar bq)_{V-A}\sum_{q'=u,d,s,c}(\bar q'q')_{V+A}\,,\,\,\,
Q_6^{(q)} = (\bar b_i q_j)_{V-A}\sum_{q'=u,d,s,c}(\bar q'_j q'_i)_{V+A}& &\label{Q6}
\eea
and 
\bea\label{Q7}
Q_7^{(q)} = \frac32(\bar bq)_{V-A}\sum_{q'}e_{q'}(\bar q'q')_{V+A}\,,\,\,\,
Q_8^{(q)} = \frac32(\bar b_i q_j)_{V-A}\sum_{q'}e_{q'}(\bar q'_j q'_i)_{V+A}& &\\
Q_9^{(q)} = \frac32(\bar bq)_{V-A}\sum_{q'}e_{q'}(\bar q'q')_{V-A}\,,\,\,\,
Q_{10}^{(q)} = \frac32(\bar b_i q_j)_{V-A}\sum_{q'}e_{q'}(\bar q'_j q'_i)_{V-A}\,.& &
\label{Q10}
\eea
Each term in the Hamiltonian (\ref{Ham}) contains a product 
$\bar q\bar q q$ which transforms as $\bar{\bf 3}\otimes \bar{\bf 3}\otimes {\bf 3} =
\overline{\bf 15} \oplus {\bf 6}\oplus \bar{\bf 3}\oplus \bar{\bf 3} $ under
flavour SU(3). When expressed in terms of well-defined SU(3) transformation properties,
the tree part of the Hamiltonian (\ref{Ham}) reads (without a factor of $G_F/\sqrt2$)
\begin{eqnarray}\label{T}
{\cal H}_T = \lambda_u^{(s)}
[\frac12(c_1-c_2)(-\bar {\bf 3}^{(a)}_{I=0} - {\bf 6}_{I=1}) +
\frac12(c_1+c_2)(-\overline {\bf 15}_{I=1} - \frac{1}{\sqrt2}\overline {\bf 15}_{I=0}
+\frac{1}{\sqrt2}\bar {\bf 3}^{(s)}_{I=0})] & &\nonumber\\
+ \lambda_u^{(d)}
[\frac12(c_1-c_2)({\bf 6}_{I=\frac12} - \bar {\bf 3}^{(a)}_{I=\frac12}) +
\frac12(c_1+c_2)(-\frac{2}{\sqrt3}\overline {\bf 15}_{I=\frac32} - 
\frac{1}{\sqrt6}\overline {\bf 15}_{I=\frac12}
+\frac{1}{\sqrt2}\bar {\bf 3}^{(s)}_{I=\frac12})] ~.& &
\end{eqnarray}
We denoted here the combinations of CKM factors $\lambda_{q'}^{(q)}=V^*_{q'b}V_{q'q}$.
There are two $\bar{\bf 3}$ operators, which were chosen to be symmetric, respectively
antisymmetric under permutations of the two $q$ fields in $qq\bar q$. The explicit
form of the operators in (\ref{T}) can be found in \cite{GPY}.

The final state in the decay consists of two octet pseudoscalar Goldstone bosons.
Bose symmetry constrains its flavour wave function to be symmetric,
which allows only certain representations
$[{\bf 8}\otimes {\bf 8}]_S = {\bf 27} \oplus {\bf 8} \oplus  {\bf 1}$.
The most general form of the decay matrix element induced by the Hamiltonian
(\ref{T}) is given by the Wigner-Eckart theorem, which is written in tensor 
language as (omitting the CKM factors)
\bea\label{WE}
{\cal H} &=& 
\langle {\bf 27}|\!| \overline{{\bf 15}} |\!|{\bf 3}\rangle
\bar M^{i_1 i_2}_{j_1 j_2} H^{j_1}_{i_1 i_2} B^{j_2} + 
\langle {\bf 8}|\!| \overline{{\bf 15}} |\!|{\bf 3}\rangle
\bar M^{i_1}_{j_1} H^{j_1}_{i_1 i_2} B^{i_2}\\
& &\hspace{-1cm} + \langle {\bf 8}|\!| {\bf 6} |\!|{\bf 3}\rangle
\epsilon_{abc} \bar M^{a}_{i} H^{ib} B^{c}
+ \langle {\bf 8}|\!| {\bar {\bf 3}}^{(a)} |\!|{\bf 3}\rangle
\bar M^{i}_{j} H_{i} B^{j}
+ \langle {\bf 1}|\!| {\bar{\bf 3}}^{(a)} |\!|{\bf 3}\rangle
\bar M H_{i} B^{j}\,.\nonumber
\eea
We denoted here with $M$ the possible tensors which can be formed from the usual
matrix of octet pseudoscalar $P^i_j = 1/\sqrt2 \pi^a\lambda^a$, corresponding to 
the mentioned symmetric representations of SU(3), and with $H$ tensors
appearing in the SU(3) decomposition of the weak Hamiltonian.
The expansion of (\ref{WE}) gives any $B$ decay
amplitude into two pseudoscalars as a linear combinations of reduced SU(3) matrix 
elements. The results are tabulated in an easy to use form in the Appendix of \cite{SU(3)3}.

There exists an equivalent description of SU(3) amplitudes in terms of quark diagrams
\cite{GHLR}, wherein a decay amplitude is decomposed into contributions which can be
associated with certain quark topologies. There are six graphical amplitudes, denoted
with $T$ (tree), $C$ (color-suppressed), $A$ (annihilation), $E$ ($W$-exchange), 
$P$ (penguin) and $PA$ (penguin annihilation). Factorization approximation combined
with quark models for the form-factors can be used to determine these graphical
amplitudes (see, e.g. \cite{fact,fact1,fact2}). In this way a hierarchy emerges, 
according to which 
the dominant amplitude is $T$, followed by $C$ which is smaller by a factor
$a_2/a_1\simeq 0.2$. 
The annihilation-type amplitudes $A$ and $E$ are predicted to be further suppressed by 
a factor $f_B/m_B\simeq 0.05$ relative to $T$ (they can be however enhanced by rescattering
effects \cite{rescatt1,rescatt1.5,Uspin1,rescatt2,rescatt3,AtSo,Uspin2,Nr,He}). 
The QCD penguin 
amplitude $P$ contributes to 
$\Delta S=0$ decays at the same order as $C$, and the $PA$ amplitude is suppressed relative 
to it as in the
case of $A$ and $E$. This additional dynamical information makes the 
graphical method more predictive than the group-theoretical approach discussed above.

We quote for later use the $B^+\to K\pi$ decay amplitudes in
quark diagram language.

\bea\label{A(K0pi+)}
& &A(B^+\to  K^0\pi^+) =\\
& &\qquad  \lambda_u^{(s)}(P_u+A) + \lambda_c^{(s)}P_c +
\lambda_t^{(s)}(P_t+P^{EW}_t(B^+\to  K^0\pi^+))~,\nonumber\\
& &\sqrt2 A(B^+\to  K^+\pi^0) =\\
& &\quad  -\lambda_u^{(s)}(T+C+P_u+A) - \lambda_c^{(s)}P_c +
\lambda_t^{(s)}(-P_t+\sqrt2 P^{EW}_t(B^+\to  K^+\pi^-))~.\nonumber
\end{eqnarray}
The unitarity of the CKM matrix can be used to eliminate the charm penguin term 
$P_c$ with the help of the relation $\lambda_c^{(s)} = -\lambda_u^{(s)}-\lambda_t^{(s)}$
by absorbing it into $P_{uc}\equiv P_u-P_c$ and $P_{tc}\equiv P_t-P_c$.

\subsection{U-spin symmetry}

At the first sight, the weak Hamiltonian (\ref{T}) appears to contain all
possible SU(3) representations allowed by the quark structure of the four-quark
operators, which would imply that no special symmetry relations exist among
decay amplitudes. In fact, an examination of the quark content of the Hamiltonian 
(\ref{Ham}) shows that it transforms as a doublet 
under $U$-spin symmetry (the subgroup of SU(3) exchanging $d$ and $s$ quarks).
Although the $\overline{\bf 15}$ representation contains both $U=1/2,3/2$ components, 
the $U=3/2$ piece cancels in the specific combinations $\overline{\bf 15}_{I=1}
+ \frac{1}{\sqrt2}\overline{\bf 15}_{I=0}$ and
$\overline{\bf 15}_{I=3/2}+\frac{1}{2\sqrt2} \overline{\bf 15}_{I=1/2}$
appearing in (\ref{T}).
The most useful amplitude relations to be used in the following are consequences of 
this symmetry property. 

The weak Hamiltonian (\ref{T}) can be written as
\bea\label{HamUspin}
{\cal H}_W &=& \left( V_{ub}^* V_{ud} {\cal T}^{(-\frac12)} +
V_{tb}^* V_{td} {\cal P}^{(-\frac12)}\right) - 
\left( V_{ub}^* V_{us} {\cal T}^{(+\frac12)} + 
V_{tb}^* V_{ts} {\cal P}^{(+\frac12)}\right)\,,
\eea
with ${\cal T}^{(U_3)}$ and ${\cal P}^{(U_3)}$ two $U=1/2$ operators standing for ``tree'' 
and ``penguin'' contributions respectively. The latter includes both strong and electroweak
penguin operators.

From the point of view of $U$-spin symmetry, the octet of pseudoscalar Goldstone
bosons contains one $U$-spin triplet ${\cal U}_1$, two doublets ${\cal U}_2$,
${\cal U}_3$ and one singlet ${\cal U}_4$. Their components are
\bea
{\cal U}_1 = 
\left( \begin{array}{c}
K^0 \\
\frac{\sqrt3}{2}\eta_8 - \frac12\pi^0 \\
-\bar K^0 \end{array}\right)\,,\quad
{\cal U}_2 = 
\left( \begin{array}{c}
K^+ \\
-\pi^+ \end{array}\right)\,,\quad
{\cal U}_3 = 
\left( \begin{array}{c}
\pi^- \\
-K^- \end{array}\right)
\eea
and ${\cal U}_4 = \frac{\sqrt3}{2}\pi^0 + \frac12\eta_8$.

To demonstrate the power of $U$-spin symmetry we derive a triangle relation \cite{GRL}
connecting the ``tree'' contributions to the $\Delta S=1$ and $\Delta = 0$ $B^+$ decays
\beq\label{triangle}
A(B^+\to K^0\pi^+) + \sqrt2 A(B^+\to K^+\pi^0) = \frac{V_{us}}{V_{ud}}
\sqrt2 A(B^+\to \pi^+\pi^0)\,.
\eeq
This relation (more precisely its extension including EWP contributions)
plays an important role in certain methods of bounding \cite{NR1} or determining 
\cite{NR2,BuFl,GPY,N} the weak phase $\gamma$ from $B\to K\pi$ decays.
The strong penguin component in ${\cal P}$ does not contribute to either side of this
relation because of isospin constraints. However, the electroweak penguin components with 
$I=1$ and $I=3/2$ respectively do contribute \cite{DesHe,Fl}, which will introduce a 
correction to Eq.~(\ref{triangle}). This will be discussed in the next section.

The final states on the left-hand side of this relation have $U_3=+1/2$ and
can be obtained by combining ${\cal U}_1\otimes {\cal U}_2$ to a total $U$-spin 1/2 
or 3/2, or by combining ${\cal U}_1\otimes {\cal U}_4$:
\bea\label{K0pi+}
& &|K^0\pi^+\rangle = -\frac{1}{\sqrt3}|[{\cal U}_1\otimes {\cal U}_2]_{\frac32}\rangle
- \sqrt{\frac23} |[{\cal U}_1\otimes {\cal U}_2]_{\frac12}\rangle\\
& &|K^+\pi^0\rangle = -\frac{1}{\sqrt6}|[{\cal U}_1\otimes {\cal U}_2]_{\frac32}\rangle
+ \frac{1}{2\sqrt3} |[{\cal U}_1\otimes {\cal U}_2]_{\frac12}\rangle
+ \frac{\sqrt3}{2} |[{\cal U}_1\otimes {\cal U}_4]_{\frac12}\rangle\\
& &|K^+\eta_8\rangle = \frac{1}{\sqrt2}|[{\cal U}_1\otimes {\cal U}_2]_{\frac32}\rangle
- \frac{1}{2} |[{\cal U}_1\otimes {\cal U}_2]_{\frac12}\rangle
+ \frac{1}{2} |[{\cal U}_1\otimes {\cal U}_4]_{\frac12}\rangle\,.
\eea
In the strangeless sector one has the $U_3=-1/2$ states
\bea
& &|\pi^0\pi^+\rangle = \frac{1}{\sqrt6}|[{\cal U}_1\otimes {\cal U}_2]_{\frac32}\rangle
+ \frac{1}{2\sqrt3} |[{\cal U}_1\otimes {\cal U}_2]_{\frac12}\rangle -\frac{\sqrt3}{2}
|[{\cal U}_1\otimes {\cal U}_4]_{\frac12}
\rangle\\\label{K+K0bar}
& &|K^+\bar K^0\rangle = -\frac{1}{\sqrt3}|[{\cal U}_1\otimes {\cal U}_2]_{\frac32}\rangle
+ \sqrt{\frac23} |[{\cal U}_1\otimes {\cal U}_2]_{\frac12}\rangle\\
& &|\pi^+\eta_8\rangle = -\frac{1}{\sqrt2}|[{\cal U}_1\otimes {\cal U}_2]_{\frac32}\rangle
- \frac12 |[{\cal U}_1\otimes {\cal U}_2]_{\frac12}\rangle
- \frac12 |[{\cal U}_1\otimes {\cal U}_4]_{\frac12} \,.
\eea
The initial state in $B^+$ decays is a $U$-spin singlet. Using the above expressions
for the final states, the relation (\ref{triangle}) follows simply as a consequence
of the absence of a $U=3/2$ term in the weak Hamiltonian.

Another useful application of the $U$-spin symmetry consists in the existence of pairs of
processes which are described by the same strong amplitudes, multiplied with different
CKM factors. This is the case, e.g. with the $B^+$ decays into $K^0\pi^+$ (\ref{K0pi+}) 
and $K^+\bar K^0$ (\ref{K+K0bar}), for which the final states contain the same $U=1/2$
$U$-spin multiplet. This gives \cite{Uspin1,Uspin2}
\bea\label{U1}
A(B^+\to K^0\pi^+) &=& V_{ub}^* V_{us} A + V_{tb}^* V_{ts} P\\\label{U2}
A(B^+\to K^+\bar K^0) &=& V_{ub}^* V_{ud} A + V_{tb}^* V_{td} P\,,
\eea
with $A$ and $P$ the reduced matrix elements of the operators ${\cal T}$ and ${\cal P}$
in (\ref{HamUspin}). Knowledge of the ratio of charge-averaged
rates for such a pair can be used to constrain the ratio of strong amplitudes entering both 
of them 
\beq\label{Ubound}
|A/P| < \lambda\sqrt{\frac{B(B^\pm\to K^0\pi^\pm)}{B(B^\pm\to K^\pm\bar K^0)}}\,.
\eeq
Also, the CP asymmetries of two such processes are equal and of opposite sign \cite{Uspin2}.
Similar relations have been used for the pair of decay amplitudes
$(B^0, B_s)\to J/\psi K_S$ \cite{FlJpsi}
and for $A(B^0\to \pi^+\pi^-)$ and $A(B_s\to K^+ K^-)$
\cite{bound,FlBs}.

\section{Electroweak penguin effects}

The contributions of the electroweak penguin operators $Q_{7-10}$ (\ref{Q7})-(\ref{Q10}) are 
suppressed relative to those of the strong penguins $Q_{3-6}$ (\ref{Q3})-(\ref{Q6})
by roughly a factor of $\alpha_{\rm e.m.}/(\alpha_s\sin^2\theta_W) \simeq 0.17$
\cite{GHLREWP,FlReview}, which is not negligibly small.
They are especially significant in penguin-dominated decays like $B\to K\pi$, where
the magnitude of the EWP amplitudes is comparable to that of the tree amplitudes. 
Therefore, a precise control over their effects is important for an understanding of
these decays.

The Wilson coefficients $c_{7-10}$ have been computed to next-to-leading order (see
\cite{NLO} for a review) with the results (at the $m_b$ scale)
\beq
(c_7,\,c_8,\,c_9,\,c_{10})\,=\,(-0.002,
0.054, -1.292, 0.263)\alpha_{\rm e.m.}\,.
\eeq
Neglecting the small contributions of the operators $c_{7,8}$ leads to important
simplifications \cite{Nr,Flr}, as the remaining EWP operators $Q_{9,10}$ are related 
by a Fierz
transformation to the current-current operators $Q_{1,2}$. Performing a SU(3) decomposition
one obtains the following expression for the EWP Hamiltonian in terms of the
$(V-A)\times (V-A)$ operators introduced in (\ref{T})
\begin{eqnarray}\label{EWP}
{\cal H}_{EWP} &\simeq & \frac{G_F}{\sqrt2}
\left\{-\lambda_t^{(s)}\left(c_9 Q^{(s)}_9 + c_{10} 
Q^{(s)}_{10}\right) -
\lambda_t^{(d)}\left(c_9 Q^{(d)}_9 + c_{10} Q^{(d)}_{10}\right)\right\} = \\
& &\hspace{-1cm} \frac{G_F}{\sqrt2}\left\{
-\frac{\lambda_t^{(s)}}{2}\left(
\frac{c_9-c_{10}}{2}(3\cdot {\bf 6}_{I=1} + \bar {\bf 3}^{(a)}_{I=0} ) +
\frac{c_9+c_{10}}{2}( -3\cdot\overline {\bf 15}_{I=1} 
-\frac{3}{\sqrt2}\overline {\bf 15}_{I=0}
-\frac{1}{\sqrt2}\bar {\bf 3}^{(s)}_{I=0} )\right)\right.\nonumber\\
& &\hspace{-2cm}\left. -\frac{\lambda_t^{(d)}}{2}\left(
\frac{c_9-c_{10}}{2}(-3\cdot {\bf 6}_{I=\frac12} + \bar {\bf 3}^{(a)}_{I=\frac12} ) +
\frac{c_9+c_{10}}{2}( -\sqrt{\frac32}\cdot\overline {\bf 15}_{I=\frac12} 
-2\sqrt3\cdot \overline {\bf 15}_{I=\frac32}
-\frac{1}{\sqrt2}\bar {\bf 3}^{(s)}_{I=\frac12} )\right)\right\}~.\nonumber
\end{eqnarray}
Now the SU(3) methods discussed in Sec.~2 can be applied to express the EWP amplitude
corresponding to any $B$ decay in terms of ``tree'' amplitudes alone. The results
have been presented in \cite{GPY} in a quark diagram language, which has the advantage
of allowing an immediate insight into the relative size of different contributions.
In particular, this justifies the color-suppression of certain EWP contributions conjectured 
in \cite{GHLREWP,Flr}.

The $U$-spin formalism discussed in Sec.~2.1 can be used to give a simple derivation
of the correction to
the triangle relation (\ref{triangle}) arising from EWP effects \cite{NR1}.
As mentioned, these corrections appear because the EWP Hamiltonian contains
$I=1$ operators in the $\Delta S=1$ sector and $I=3/2$ operators in the $\Delta S=0$ 
sector respectively, as one can see from (\ref{EWP}). Their matrix elements can be related
thanks to the special structure of the weak Hamiltonian (\ref{HamUspin}) written in
$U$-spin symmetric form:
\bea
& &{\cal T} = 
\frac12 (c_1+c_2) {\cal D}_1 +
\frac12 (c_1-c_2) {\cal D}_2\\
& &{\cal P}^{EWP} = 
\frac12 (c_9+c_{10}) [-\frac32 {\cal D}_1 +{\cal D}_3] +
\frac12 (c_9-c_{10}) {\cal D}_4\,,
\eea
where ${\cal D}_2^{(-\frac12)}$, ${\cal D}_3^{(-\frac12)}$ and 
${\cal D}_4^{(-\frac12)}$ are $I=1/2$ operators and only ${\cal D}_1^{(-\frac12)}$ has $I=3/2$. 
This special property can be used \cite{NR1} to prove that, 
although ${\cal D}_3^{(+\frac12)}$ and ${\cal D}_4^{(+\frac12)}$ 
contain $I=1$ pieces, they do not contribute to the LHS of (\ref{triangle}). 
Therefore the EWP contribution to the LHS of (\ref{triangle}) can be expressed
solely in terms of the amplitude $A(B^+\to \pi^+\pi^0)$ induced by ${\cal D}_1$.
One obtains in this way the following generalization of (\ref{triangle}) including
the contributions of the EW penguin effects
\bea\label{triangleEWP}
& &A(B^+\to K^0\pi^+) + \sqrt2 A(B^+\to K^+\pi^0) = \\
& &\qquad 
\frac{V_{us}}{V_{ud}}\frac{f_K}{f_\pi}\sqrt2 A(B^+\to \pi^+\pi^0)
\left( 1 - \frac{c_9+c_{10}}{c_1+c_2}\frac{3}{2R_b\lambda^2}e^{-i\gamma}
\right)\,.\nonumber
\eea
In this relation one has neglected the EWP contribution to the decay amplitude
$A(B^+\to \pi^+\pi^0)$. They can be included in a model-independent way too
\cite{BuFl,GPY}, although their numerical impact turns out to be rather small, 
in accordance with earlier estimates \cite{GHLREWP}. The ratio $f_K/f_\pi\simeq 1.22$
accounts for factorizable SU(3) breaking in the leading tree amplitude.

It is interesting to note that there exist SU(3) amplitude relations which are not
affected by EWP effects. One of them has been noted by Deshpande and He \cite{DesHe},
who based on it a different method for determining $\gamma$ (see also \cite{GrRos}).
This relation follows from the absence of a $U=3/2$ term in the weak Hamiltonian:
\bea
0 &=& \langle [{\cal U}_1\otimes {\cal U}_2]_{\frac32}, U_3=+\frac12|H_W|B^+\rangle\\
 &=&
\sqrt2 A(B^+\to K^+\pi^0) - \sqrt6  A(B^+\to K^+\eta_8) + 
2 A(B^+\to K^0\pi^+)\,.\nonumber
\eea
Its analog for $\Delta S=0$ decays has been used in \cite{GP} and relates
$B^+$ decay amplitudes into strangeless final states
\bea\label{GP}
0 &=& \langle [{\cal U}_1\otimes {\cal U}_2]_{\frac32}, U_3=-\frac12|H_W|B^+\rangle\\
&=&  A(B^+\to K^+\bar K^0) + \sqrt{\frac32}  A(B^+\to \pi^+\eta_8)
- \frac{1}{\sqrt2}  A(B^+\to \pi^+\pi^0)\,.\nonumber
\eea

\section{Determining the weak phase $\gamma$ using $B\to K\pi$ decays}

A method for determining the weak phase $\gamma$ has been proposed in \cite{GRL},
requiring the $B^+$ decay rates into $K^0\pi^+, K^+\pi^0, \pi^+\pi^0$ and their charge 
conjugates. This method, subsequently improved in \cite{NR2} by including EW penguin 
effects, rests on the following assumptions:

a) flavor SU(3) symmetry, implied in the triangle relation (\ref{triangle}), respectively
its version (\ref{triangleEWP}) including EWP effects.

b) the absence of a term with nontrivial weak phase in the amplitude $A(B^+\to K^0\pi^+)$.
This amplitude has been given in (\ref{A(K0pi+)}) and can be rewritten as
\bea\label{epsA}
A(B^+\to K^0\pi^+) = -A\lambda^2 P\left(1 + \varepsilon_A e^{i\phi_A} e^{i\gamma}
\right)
\eea
with $\varepsilon_A, \phi_A$ parametrizing the magnitude and phase of the
annihilation contribution relative to the dominant penguin one. 

Neglecting the annihilation amplitude ($\varepsilon_A\simeq 0$), the SU(3) triangle 
(\ref{triangle}) and its CP conjugate can be represented together as shown in Fig.~1.
The circle has radius $\delta_{EW}$ in units of $\sqrt2 A(B^+\to \pi^+\pi^0)$, with
$\delta_{EW} = -\frac{3}{2\lambda^2 R_b}\frac{c_9+c_{10}}{c_1+c_2} = 0.66\pm 0.15$.
The relative orientation of the two triangles is fixed together with the weak phase $\gamma$
by requiring the equality of the two angles denoted $2\gamma$ in Fig.~1.

There are several sources of theoretical errors affecting this determination. First,
there are uncertainties in the value of $\delta_{EW}$ from SU(3) breaking effects and
the imprecisely known value of the ratio $|V_{ub}/V_{cb}|$. The former have been 
computed in the factorization approximation \cite{NR1,N}
and they lower $\delta_{EW}$ by $(6\pm 6)\%$
compared to its SU(3) value, although nonfactorizable SU(3) breaking, which could be
significant \cite{BuFl}, remains unknown. At present the latter dominate the error on 
$\delta_{EW}$ but they are likely to decrease as the ratio of CKM matrix elements is 
better measured.

We will focus in the following on another intrinsic uncertainty of this method, arising
from rescattering effects (assumption (b) above). As explained above, the naive 
factorization approximation suggests that the component with weak phase $\gamma$ in the
amplitude (\ref{epsA}) is suppressed by a factor $f_B/m_B\simeq 0.05$ and is thus
negligibly small. However, dynamical calculations
\cite{rescatt1,rescatt1.5,rescatt2,rescatt3,rescatt4}
suggest that rescattering effects can induce a nonnegligible value for $\varepsilon_A$.
For example, elastic rescattering through a color-allowed intermediate
state as in $B^+\to \{ K^+\pi^0\}\to K^+\pi^0$ can conceivably enhance the annihilation
contribution.

The $U$-spin relation (\ref{Ubound}) can be used to give an upper bound on the magnitude
of these effects $\varepsilon_A < 0.18$, which is not yet very stringent.
We used here the CLEO results \cite{Frank} $B(B^\pm\to K^\pm K^0)<0.9\cdot 10^{-5}$
(at 90\% CL) and $B(B^\pm\to K^0\pi^\pm)=(1.4\pm 0.5\pm 0.2)\cdot 10^{-5}$.
We will adopt in our following estimates the value $\varepsilon_A=0.1$.

The complete set of $B^+\to K\pi$ decay amplitudes is defined by specifying $\varepsilon,
\phi_P, \gamma$ and the rescattering parameters $\varepsilon_A, \phi_A$, where we
denote the ``tree-to-penguin'' ratio
\beq
\varepsilon = \lambda\frac{f_K}{f_\pi}\sqrt{\frac{B(B^\pm\to \pi^\pm\pi^0)}{B(B^\pm\to
K^0\pi^\pm)}}
\eeq
and the relative phase $\phi_P=$Arg$(P/(T+C))$. One can simulate sets of decay amplitudes
corresponding to given values of these parameters and study the effects of the rescattering
effects on the extracted value of $\gamma$.

In Fig.~2 are shown the results of such a simulation using the input values 
$\varepsilon=0.24$, $\varepsilon_A=0.1$ and $\gamma=76^\circ$. In Fig.~2(a) the
output value of $\gamma$ is plotted as a function of $\phi_A$ at $\phi_P=60^\circ$ and
$\phi_P=90^\circ$, and in Fig.~2(b) the dependence of $\gamma$ is shown as function
of $\phi_P$ at $\phi_A=0^\circ$.
The most notable feature of these results is the large deviation of the extracted $\gamma$ 
from its physical value for a strong phase $\phi_P$ around $90^\circ$, of about 14$^\circ$.
This example illustrates the possible significance of the rescattering corrections on this
method, even for moderate values of $\varepsilon_A\simeq 0.1$. 

A modified version of this method for determining $\gamma$ has been proposed in \cite{N},
with the view of minimizing the rescattering effects. This method is formulated in terms
of two quantities $R_*$ and $\tilde A$ defined by
\bea
& &R_* \equiv \frac{B(B^\pm\to K^0\pi^\pm)}{2B(B^\pm\to K^\pm\pi^0)}~,
\\
& &\tilde A \equiv \frac{B(B^+\to K^+\pi^0)-B(B^-\to K^-\pi^0)}{B(B^\pm\to 
K^0\pi^\pm)} - \frac{B(B^+\to K^0\pi^+)-B(B^-\to \bar K^0\pi^-)}{2B(B^\pm\to 
K^0\pi^\pm)}~.\nonumber
\eea
These quantities do not contain ${\cal O}(\varepsilon_A)$ terms; their dependence 
on the rescattering parameter $\varepsilon_A$ appears only at order 
${\cal O}(\varepsilon\varepsilon_A)$. Therefore, it was argued in \cite{N}, 
the determination of $\gamma$, by setting $\epsilon_A=0$ in the expressions for 
$R_*$ and $\tilde A$, is insensitive to rescattering effects. 
This procedure gives two equations for $\gamma$ and $\phi$ which can 
be solved simultaneously from $R_*$ and $\tilde A$. Using two pairs of input 
values for ($R_*, \tilde A$) (corresponding to a restricted range for $\phi_A$ 
and $\phi_P$) seemed to indicate that the error in $\gamma$
for $\varepsilon_A=0.08$ is only about $5^{\circ}$. 

In Fig.~3 are shown the results of such an analysis carried out for the entire 
parameter space of $\phi_A$ and $\phi_P$.  
Whereas the angle $\phi_P$ can
be recovered with small errors, the results for $\gamma$ show the same 
large rescattering effects for values of $\phi_P$ around 90$^\circ$ as in Fig.~2.
(A slight improvement is the absence of a discrete ambiguity in the value of 
$\gamma$.) These results indicate that the large deviation of $\gamma$ from its 
physical value for $\phi_P=90^\circ$ is a general phenomenon, common to all 
variants of this methods. Some information about the size of the
expected error can be obtained by first determining $\phi_P$. Values not too 
close to 90$^\circ$ would be an indication for a small error.

\subsection{Eliminating the rescattering effects using additional processes}

Several modifications of the method discussed above have been proposed 
\cite{Flr,BuFl,GP,AgDe}, which use additional processes in order to completely 
eliminate the rescattering contributions. All of these methods make use of the
decays $B^\pm\to K^\pm K^0$ which are related by $U$-spin to the amplitude
$B^\pm\to K^0\pi^\pm$ affected by rescattering by (\ref{U1}), (\ref{U2}). Using these
relations one can see that the rescattering effects cancel out in the difference
$A(B^+\to K^0\pi^+) - \lambda A(B^+\to K^+\bar K^0)$.

This is illustrated in Fig.~4, where in addition to the SU(3) triangles of Fig.~3
the amplitudes $\lambda A(B^+\to K^+\bar K^0)$ and of its CP conjugate are shown as
the segments $OC$ and $OD$ respectively. Assuming that the positions of $OC$ and $OD$ 
are known, then the relative orientation of the $B\to K\pi$ triangles and thereby $\gamma$
can be fixed by requring the equality of the two angles marked $2\gamma$ in Fig.~4.
The various existing methods in the literature differ in the way the positions of
the $OC$ and $OD$ segments are determined.

A minimal extension has been proposed in \cite{Flr,BuFl} which requires, in addition to 
$B^+\to K\pi, \pi^+\pi^0$ data, only the charge-averaged rate for $B^\pm\to K^\pm K^0$.
In the geometrical formulation given in \cite{NR2}, this method works
by requiring the equality of the two segments $|YC|=|YD|$ (both considered as functions
of $\gamma$) in Fig.~4. Due to the smallness of the rescattering contribution relative to
the penguin amplitude, this equality is almost automatical for most values of $\gamma$,
which is to say that small errors in the amplitudes $A(B^+\to K^+\bar K^0)$ are amplified
in the extracted value of $\gamma$. Also, SU(3) breaking effects introduce large errors, 
which can be however controlled if the direct CP asymmetry of the $K^+\bar K^0$ mode is 
measured.

An improvement of this approach has been given in \cite{GP}, where the positions of the
segments $OC$ and $OD$ are determined independently of the $B^+\to K\pi$ data, with the
help of the SU(3) relation (\ref{GP}). The uncertainty in the position of the point $Y$
due to SU(3) breaking is naturally small because the sides $OC$ and $OD$ themselves are
small, relative to the long side of the triangle (\ref{GP}). Naive dimensional estimates
\cite{GP} suggest that the SU(3) breaking-induced error on $\gamma$ is of
the order of a few degrees, which is confirmed by a detailed numerical study \cite{GP2}.
One additional problem with this method is introduced by the $\eta-\eta'$ mixing, whose 
treatment will add some model dependence. This can be avoided by using instead an alternative 
approach using $B^0\to K\pi$ and $B_s$ decays \cite{GP}. However, it remains to be seen if 
the statistical 
errors due to the necessity of combining nine different decay rates will not overweigh the 
theoretical advantages of this method.

\section{Conclusions}

Nonleptonic weak decays of the $B$ mesons are a valuable source of information about the
elements of the CKM matrix. In particular, the penguin-dominated decays $B\to K\pi$ can provide
useful constraints \cite{FM,GRbound,NR1} and determinations \cite{GRL,NR2,Flr,BuFl,GPY}
of the weak phase $\gamma$, which complement those from global fits of the unitarity triangle. 
Although the focus of this talk has been on charged $B$
decays, useful information can be obtained also by combining $B^0$ with $B^+$ decay
data \cite{Fl,FM,GRbound,Flr,BuFl}. 

While the electroweak penguin contributions to the 
determination of $\gamma$ from $B^+\to K\pi, \pi^+\pi^0$ decays 
can be controlled using SU(3) symmetry, the rescattering effects are potentially
significant. Depending on the precise value of a strong phase $\phi_P$ (which can be
determined fairly precisely), the corresponding uncertainty on $\gamma$ can be as 
large as $\pm 15^\circ$. Several methods exist which make it possible to completely
eliminate these effects with the help of additional decays.

\acknowledgements

It is a pleasure to thank Michael Gronau and Tung-Mow Yan for discussions and collaboration
on the subjects discussed in this talk. This work has been supported by the National
Science Foundation.


\newpage

\begin{figure}[hhh]
 \begin{center}
 \mbox{\epsfig{file=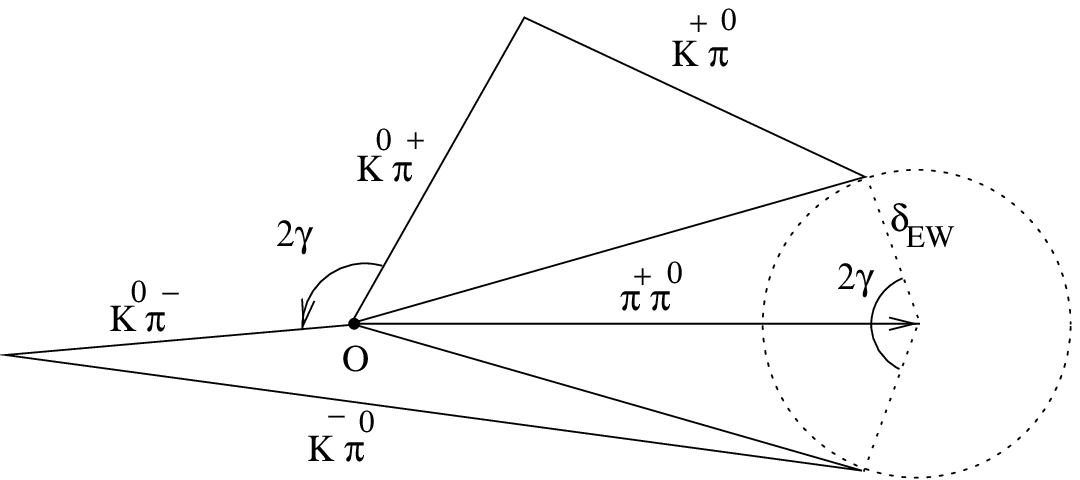,width=10cm}}
 \end{center}
 \caption{
Graphical representation of the SU(3) amplitude relation (\ref{triangleEWP}) and of
its CP conjugate used in the determination of the weak angle $\gamma$.}
\label{fig1}
\end{figure}

\begin{figure}[hhh]
 \begin{center}
 \mbox{\epsfig{file=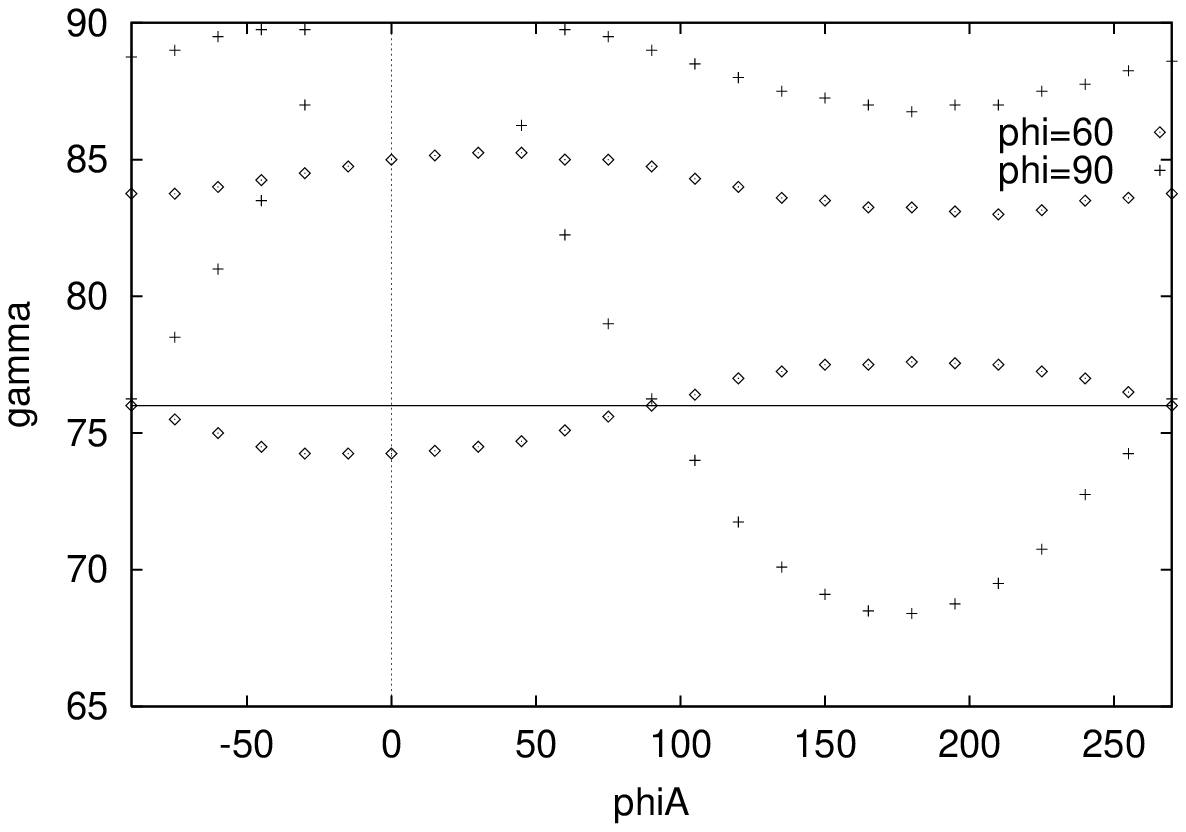,width=7cm}}
 \mbox{\epsfig{file=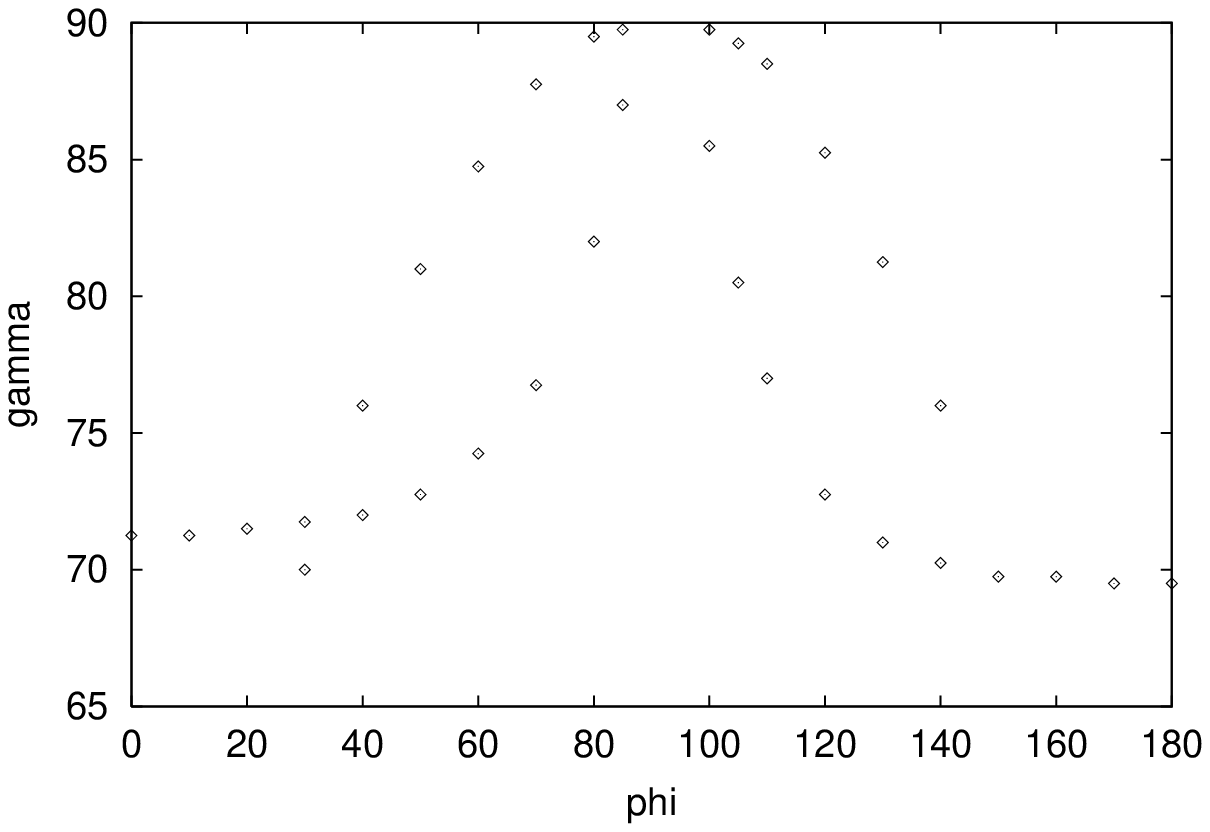,width=7cm}}\\
\hspace*{5mm}\mbox{(a)}\hspace*{7cm}\mbox{(b)}
 \end{center}
 \caption{Rescattering effects on the determination of the weak phase $\gamma$ from
$B^+\to K\pi$ decays. 
(a) - the dependence of $\gamma$ on $\phi_A$,
for two values of $\phi_P=60^\circ$ and $\phi_P=90^\circ$; (b) - the dependence
of the solution on $\phi_P$, for $\phi_A=0^\circ$. (both graphs correspond to
$\varepsilon_A=0.1, \gamma_{\rm phys}=76^\circ$) }
\label{fig2}
\end{figure}

\begin{figure}[hhh]
 \begin{center}
 \mbox{\epsfig{file=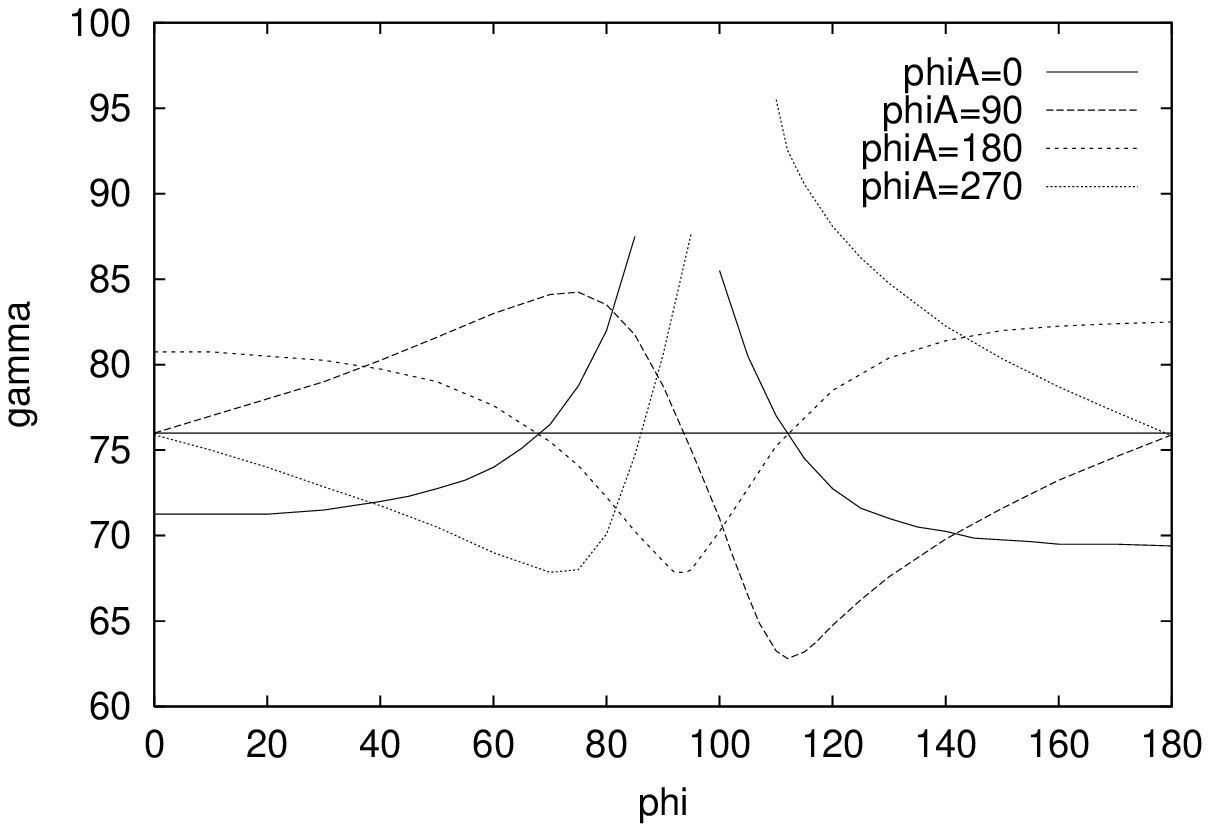,width=7cm}}
 \mbox{\epsfig{file=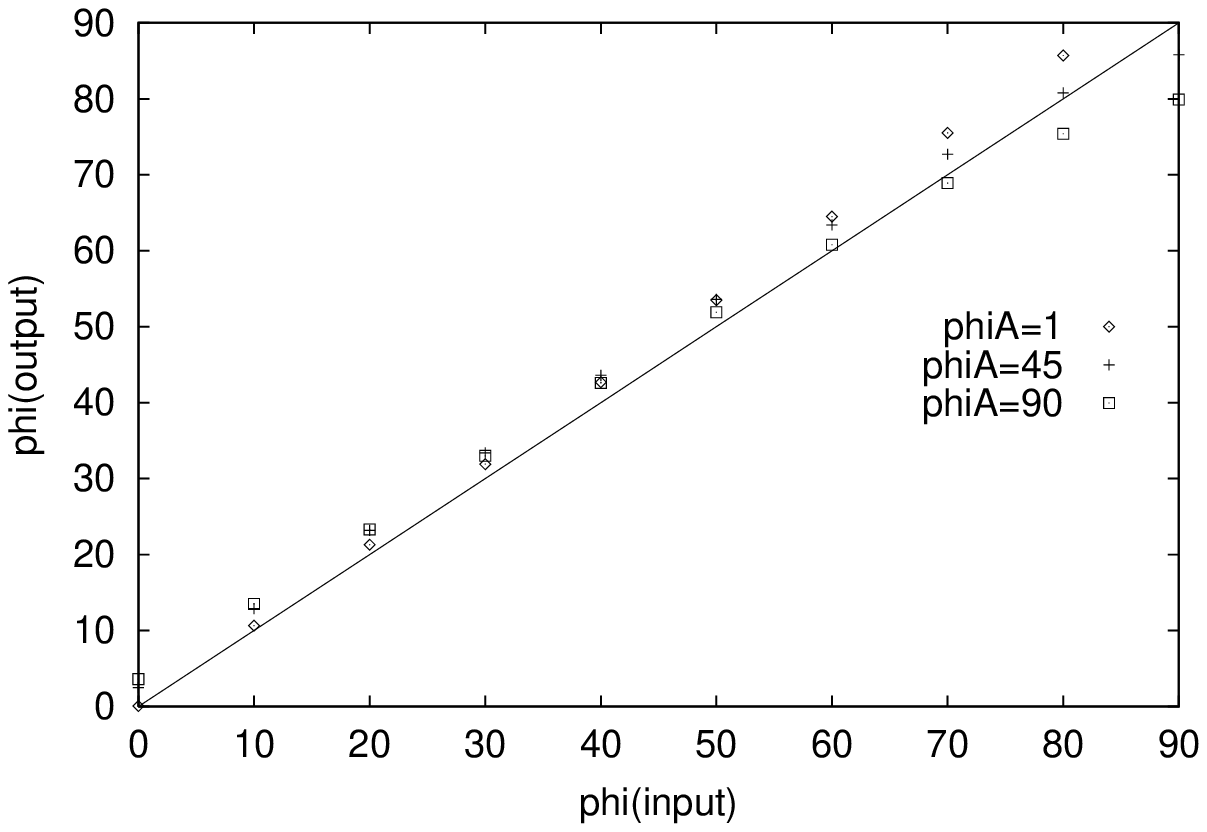,width=7cm}}\\
\hspace*{5mm}\mbox{(a)}\hspace*{7cm}\mbox{(b)}
 \end{center}
 \caption{(a) - the weak phase $\gamma$ extracted from the method using
the parameters $(R_*,\tilde A)$, as a function of the strong phase 
$\phi_P$ for several values of $\phi_A$ ($\varepsilon_A=0.1$). 
The horizontal line shows the assumed physical value of $\gamma=76^\circ$.
(b) - the strong phase $\phi_P$ can be reconstructed using the $(R_*,\tilde A)$ data.}
\label{fig3}
\end{figure}

\begin{figure}[hhh]
 \begin{center}
 \mbox{\epsfig{file=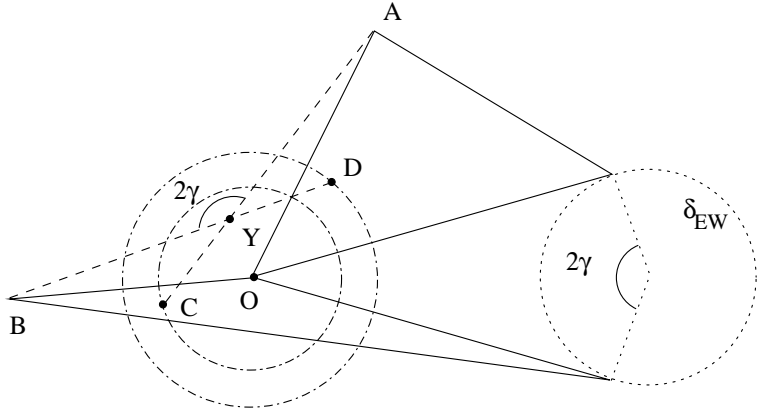,width=10cm}}
 \end{center}
 \caption{Eliminating rescattering corrections with the help of $B^\pm\to K^\pm K^0$ decays. 
The lines $OC$ and $OD$ denote the amplitudes $\lambda A(B^\pm\to K^\pm K^0)$.
}
\label{fig4}
\end{figure}


\end{document}